\documentstyle[twoside,fleqn,espcrc2,epsf]{article}

\newcommand{\bra}[1]{\left< #1 \right|}

\newcommand{\ket}[1]{\left| #1 \right>}

\title{Latest Results from the SGO Collaboration}
\author{Presented by S.~Collins\address{Dept. of Physics and Astronomy, Glasgow University, Glasgow, G12 8QQ, Scotland}
\thanks{In collaboration with A.~Ali~Khan 
and C.~T.~H.~Davies, Glasgow Univ.; 
J.~Shigemitsu, Ohio State Univ.; 
U.~M.~Heller and J.~H.~Sloan, SCRI at Florida State Univ..}}
       
\begin{document}

\begin{abstract}
We present results for the spectrum and decay constants of B mesons
from NRQCD using dynamical configurations at $\beta=5.6$ with two
flavours of staggered fermions. The light quarks are generated using
the Clover action with tadpole improvement. 
\end{abstract}

\maketitle
\section{INTRODUCTION}
We present a study of the heavy-light meson spectrum and decay
constants, with a focus on investigating heavy quark symmetry as well
as providing quantitative predictions.  Our approach is to use NRQCD
for the heavy quark and in this analysis the action includes terms to
$O(1/M_0)$:
\label{intro}
\begin{eqnarray}
{\cal L}_{NR} & = &  {\cal L}_{stat} + ({\cal L}_{kin}+ {\cal L}_{hyp})/2M_0\\
{\cal L}_{stat} & = & Q^\dagger D_t Q\\
{\cal L}_{kin} & = & -Q^\dagger \vec{D}^2 Q\\
{\cal L}_{hyp} & = & -Q^\dagger c_B \vec{\sigma}\cdot \vec{B}Q.
\end{eqnarray}
Matrix elements must be calculated consistently to this order. Here,
we restrict the analysis to the tree-level operators for the axial and
vector current to $O(1/M_0)$~\cite{spect},
\begin{eqnarray}
{\cal O}_1 = \bar{q} \Gamma Q\hspace{1cm} {\cal O}_2 = \bar{q} \Gamma\gamma_iD_iQ /2M_0
\end{eqnarray}
where $\Gamma=\gamma_5\gamma_0$ and $\gamma_i$ respectively. Tadpole
improvement is used throughout and then $c_B$ is given the tree-level value
of $1.0$.

The heavy quark propagators were generated using the evolution equation
\begin{eqnarray}
G_{t+1} & = & (1-\frac{a\delta H}{2})\left(1-\frac{aH_0}{2n}\right)^n 
U_4^{\dagger} \nonumber\\ 
&  & \times\left(1-\frac{aH_0}{2n}\right)^n 
(1-\frac{a\delta H}{2})G_t \label{evol}
\end{eqnarray}
for all timeslices. Equation~(\ref{evol}) is fully consistent to
$O(1/M_0)$. This differs from the evolution equation previously used
in~\cite{spect} by a factor of $1{-}\frac{a\delta H}{2}$ on
the source timeslice, and our results had a
systematic error of $O(a\Lambda_{QCD}/M)$. However, we found this
error to be small~\cite{spect} and our previous findings are not
affected.

The simulation was performed on 100 $16^3\times 32$ lattices at
$\beta=5.6$ with two flavours of dynamical staggered fermions,
generated by the HEMCGC collaboration. The light quark propagators
were computed using the Clover action with tadpole improvement, at
three masses of light quark, $\kappa_l=0.1385$, $0.1393$ and $0.1401$.
The heavy quark propagators were computed over a wide range of bare
quark mass, $aM_0=0.8{-}10.0$, in order to study heavy quark
symmetry. Full details of our analysis can be found in~\cite{spect}.
\section{RESULTS}	
\begin{figure}
\setlength{\unitlength}{.017in}
\begin{picture}(80,120)(0,930)
\put(15,940){\line(0,1){100}}
\multiput(13,950)(0,50){2}{\line(1,0){4}}
\multiput(14,950)(0,10){10}{\line(1,0){2}}
\put(12,950){\makebox(0,0)[r]{5.3}}
\put(12,1000){\makebox(0,0)[r]{5.8}}
\put(12,1030){\makebox(0,0)[r]{ }}
\put(12,1040){\makebox(0,0)[r]{GeV}}

\put(50,940){\makebox(0,0)[t]{$B$}}
\multiput(43,948)(3,0){5}{\line(1,0){2}}
\put(50,948){\circle*{2}}

\put(50,1014){\circle*{2}}
\put(50,1014){\line(0,1){6}}
\put(50,1014){\line(0,-1){6}}
\put(62,1020){\makebox(0,0)[t]{($2S$)}}

\put(75,940){\makebox(0,0)[t]{$B^*$}}
\multiput(68,953)(3,0){5}{\line(1,0){2}}
\put(75,951){\circle*{2}}
\put(75,951){\line(0,1){1}}
\put(75,951){\line(0,-1){1}}

\put(100,940){\makebox(0,0)[t]{$B_s$}}
\multiput(93,958)(3,0){5}{\line(1,0){2}}
\put(100,958){\circle*{2}}
\put(100,958){\line(0,1){1}}
\put(100,958){\line(0,-1){1}}

\put(125,940){\makebox(0,0)[t]{$B^{**}$}}
\multiput(118,992)(3,0){5}{\line(1,0){2}}
\multiput(118,987)(3,0){5}{\line(1,0){2}}
\put(125,996){\circle*{2}}
\put(125,996){\line(0,1){4}}
\put(125,996){\line(0,-1){4}}

\end{picture}
\caption{The errors shown are purely statistical.\label{figspect}}
\end{figure}
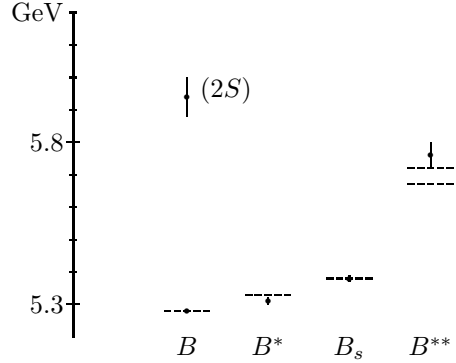

A vital part of a lattice calculation of $B$ physics is a
comprehensive calculation of the spectrum. This provides a firm
foundation for the simulation of more complicated
processes. Figure~\ref{figspect} presents the results for the low
lying $B$ spectrum; $a^{-1}=2.0$~GeV from light spectroscopy has been
used to convert to physical units and the $B$ meson mass to fix
$aM_0^b$. For the mass splittings $B_s{-}B$ and $B^{**}{-}B$, the
dominant systematic errors are estimated to be comparable to or less
than the statistical errors. With the exception of $B^*{-}B$
we find good agreement with experiment, although, the 
experimental results for the $P{-}S$ splitting are preliminary. The
discrepancy with the real world for the hyperfine splitting may be a
residual quenching effect. However, it is also possible our
value of $c_B$ may be too low by $10{-}20\%$; a nonperturbative
determination of $c_B$ is needed.

Note that the gross shape of the spectrum supports the naive picture
of a massive heavy quark with velocity $v{\sim} 0.1$ surrounded by a
light quark cloud. In this model the hyperfine spitting is expected to
be $Mv^2{\sim} 50$~MeV, while the $P{-}S$ and $2S{-}1S$ splittings are due
to excitations of the light quark and so are of 
$O(\Lambda_{QCD}){\sim} 200{-}500$~MeV. We can investigate this further by
quantifying the change in these splittings with
$M_Q$. Table~\ref{Etab} summarises the coefficients extracted from
fitting the spin-averaged ground $S$-state energy~($\overline{E}$) and
mass splittings to the function
\begin{equation}
\bar{E},\mbox{ } \Delta E = C_0 + C_1/M + C_2/M^2 +\ldots\label{fitfun}
\end{equation}
where $M$ is the $PS$ meson mass.

The results are compared with our expectations in Table~\ref{Etab}.
With the exception of the bare kinetic energy of the heavy quark
(extracted from $\bar{E}$) the overall agreement is good.  While
$\bra{1S}{\cal L}_{kin} \ket{1S}_{bare}$ is small, it is opposite in
sign to the physical result, which is positive.  This is due to our
use of tadpole-improved matrix elements~\cite{spect}; using mean field
theory the unimproved matrix element corresponds to a large positive
result, $\bra{1S}{\cal L}_{kin} \ket{1S}_{bare}^{unimproved}{\sim}+0.7$.
This latter result is in agreement with an explicit calculation by
Crisafulli et al~\cite{sach}, and demonstrates that the matrix element
is tadpole dominated. In fact, the renormalisation required to obtain
the (positive) physical quantity from the tadpole-improved
$\bra{1S}{\cal L}_{kin}
\ket{1S}_{bare}$ is small and can be calculated perturbatively.

The slope of the $2S{-}1S$ splitting gives the difference between the
physical kinetic energy of the heavy quark in the excited state and
ground state mesons; it is a physical quantity since the
renormalisation shift cancels in the difference. 
Note that the slope is positive, and its magnitude suggests $\bra{2S} {\cal
L}_{kin}\ket{2S}$ is large compared to $\bra{1S} {\cal
L}_{kin}\ket{1S}$.

\begin{table*}
\begin{center}
\caption{The coefficients
of various spectral quantities as determined from first order
perturbation theory in $1/M$, where $\bra{1S}$ represents an $S$-state
in the infinite mass limit. The results are in physical units~(in
powers of GeV). Note that those obtained for $2S{-}1S$ are estimates
and not obtained from a fit. $E_0^\infty$ is the energy of a heavy
quark in the static theory, calculated perturbatively.
$\kappa_l=\kappa_c$.\label{Etab}}
\begin{tabular}{|c|c|c|c|}\hline
Quantity & Coefficient & Results & Expectation \\\hline
$\overline{E}$ &  $C_0 = E^{\infty}$ & 1.00(2) & Static Result \\
               &  $\bar{\Lambda}= E^{\infty} - E_0^\infty$ & 0.3(3)  & $+\Lambda_{QCD}$\\
                      &  $C_1 = \bra{1S}{\cal L}_{kin} \ket{1S}_{bare}$  & $-0.1$(1) & $+\Lambda_{QCD}^2$\\\hline
$E(2S)-E(1S)$ & $C_0 = E^\infty(2S) - E^\infty(1S)$ & 0.5 & $+\Lambda_{QCD}$ \\
	      & $C_1 = \bra{2S} {\cal L}_{kin} \ket{2S}_{phys} - \bra{1S} {\cal L}_{kin} \ket{1S}_{phys}$ & 0.3 & $+$ve\\\hline
$E(^3S_1)- E(^1S_0)$ & $C_0=0$ & 0.000(2) &  0\\
                     & $C_1= \bra{1S} {\cal L}_{hyp} \ket{1S}_{phys}$ & 0.20(4) & $+\Lambda_{QCD}^2$ \\\hline
\end{tabular}
\end{center}
\end{table*}

Now consider a similar analysis of the vector and axial currents. We
define the $V$ and $PS$ decay constants in the same way using
\begin{eqnarray}
f\sqrt{M}  & = &   \bra{0} {\cal O}_1 \ket{P}_{NR} /\sqrt{M}\\
\delta(f\sqrt{M}) & = &  \bra{0} {\cal O}_2
\ket{P}_{NR}/\sqrt{M} \\
(f\sqrt{M})^{tot} & = & (f\sqrt{M})+ \delta(f\sqrt{M})
\end{eqnarray}
where $\delta(f\sqrt{M})$ denotes the $O(1/M)$ correction to the
current. The expansion of $f\sqrt{M}$ about the static limit is
given in HQET to $O(1/M)$~\cite{neubert} by
\begin{eqnarray}
(f\sqrt{M})^{tot} & = & (f\sqrt{M})^{stat}( 1 + c_P/M)\\
c_P & = & G_{kin} + 2dG_{hyp} + dG_{cor}/6
\label{slope}
\end{eqnarray}
where $d=3$ and ${-}1$ for $PS$ and $V$ mesons respectively. In
equation~(\ref{slope}) the slope is decomposed into the contributions
from the kinetic energy of the heavy quark, $G_{kin}$, the hyperfine
interaction, $G_{hyp}$, and the correction to the current, $G_{cor}$.
Equation~(\ref{slope}) suggests that these contributions can be
isolated by finding the slope of various combinations of the $V$ and
$PS$ decay constants.

Table~\ref{ftab} presents our results with the corresponding
contributions to the slopes, obtained using a fit function of the
form~(\ref{fitfun}). The physical quantities
$(f\sqrt{M})^{tot}_{PS}$, $(\overline{f\sqrt{M}})$ and
$(f\sqrt{M})^{tot}_{PS}/(f\sqrt{M})^{tot}_{PS}$ are naively expected
to have coefficients of the slope of $O(\Lambda_{QCD})$. However, we
find a much larger slope than this for the total decay constant.  The
slope of the spin average of the decay constants, which only depends
on $G_{kin}$, shows that it is the contribution from the kinetic
energy of the heavy quark which is large and dominates $c_P$.
Comparing with Neubert~\cite{neubert} we find good agreement for the
slopes of the physical combinations of decay constants while for the
unphysical slope of $(f\sqrt{M})_{PS}/(f\sqrt{M})_V$ and intercept of
$2M_0\delta (f\sqrt{M})_{PS}/(f\sqrt{M})_{PS}$, where agreement is not
expected, the results differ in magnitude and for the former also in
sign~\cite{spect}.

Extracting the value of the $PS$ decay constant at $M_B$ we find
\begin{equation}
Z_A^{-1}f_B = 216\pm(6) \pm(10) \pm(20)\mbox{ MeV}
\end{equation}
The first error is statistical and the second is the systematic error
arising from the clover light fermions estimated to be $O(a^2)$.  The
third is due to the omission of $O(1/M_0^2)$ terms in the NRQCD
action, and is the dominant error in $f_B$.
Using the static renormalisation factor of
$Z_A^{stat} = 0.8$ we find $f_B{\sim} 180$~MeV.

\section{CONCLUSIONS}
We have performed a comprehensive study of the heavy-light meson
spectrum and decay constants.  The question that arises from this
analysis is why there is such a large slope for the decay constant
when the deviations from the static limit for spectral quantities are
in agreement with naive expectation.  A potential model
approach~\cite{chris} suggests that this is a consistent picture and
that a large value for $c_P$ is coupled to a large value for $\bra{2S}
{\cal L}_{kin}\ket{2S}$.

\begin{table}
\begin{center}
\begin{tabular}{|c|c|c|}\hline
Quantity & $C_1/C_0$ & Results\\\hline $(f\sqrt{M})^{tot}_{PS}$ &
$G_{kin}+6G_{hyp}+G_{cor}/2$ & $-2.0$(4) \\\hline
$(\overline{f\sqrt{M}})$ & $G_{kin}$ & $-1.8$(2) \\\hline
$\frac{(f\sqrt{M})_{PS}}{(f\sqrt{M})_{V}}$ & $8G_{hyp}$ & 0.72(6)
\\\hline 
$\frac{(f\sqrt{M})^{tot}_{PS}}{(f\sqrt{M})^{tot}_{V}}$ &
$8G_{hyp}+2G_{cor}/3$ & $-0.24$ \\\hline 
Quantity & $C_0$ &
Results\\\hline $\frac{2M_0\delta(f\sqrt{M})_{PS}}{(f\sqrt{M})_{PS}}$
& $-G_{cor}$ & $-1.24$(2) \\\hline
\end{tabular}
\caption{The results are in GeV. Note that renormalisation factors 
are not included. $\kappa_l=\kappa_c$.\label{ftab}}
\end{center}
\end{table}

The authors acknowledge the support under grants from
NATO and DOE.  The computations were performed on the CM-2 at SCRI.

\end{document}